\documentstyle[11pt,epsfig]{article}

\advance \topmargin by -\headheight
\advance \topmargin by -\headsep
     
\evensidemargin \oddsidemargin
\makeatletter

\def\@sect#1#2#3#4#5#6[#7]#8{\ifnum #2>\c@secnumdepth
     \def\@svsec{}\else 
     \refstepcounter{#1}\edef\@svsec{\csname the#1\endcsname.\hskip 1em }\fi
     \@tempskipa #5\relax
      \ifdim \@tempskipa>\z@ 
        \begingroup #6\relax
          \@hangfrom{\hskip #3\relax\@svsec}{\interlinepenalty \@M #8\par}
        \endgroup
       \csname #1mark\endcsname{#7}\addcontentsline
         {toc}{#1}{\ifnum #2>\c@secnumdepth \else
                      \protect\numberline{\csname the#1\endcsname}\fi
                    #7}\else
        \def\@svsechd{#6\hskip #3\@svsec #8\csname #1mark\endcsname
                      {#7}\addcontentsline
                           {toc}{#1}{\ifnum #2>\c@secnumdepth \else
                             \protect\numberline{\csname the#1\endcsname}\fi
                       #7}}\fi
     \@xsect{#5}}

\renewcommand{\section}{\setcounter{equation}{0} \@startsection {section}{1}
   {\z@}{-3.5ex plus -1ex minus -.2ex}{2.3ex plus .2ex}{\Large\bf}}

\newcommand{\im}{\mathop{\mathrm{Im}}}

\def\IJMP #1 #2 #3 {{\it Int.\ J.\ Mod.\ Phys.}\ {\bf #1}\ (#2) #3}
\def\MPL #1 #2 #3 {{\it Mod.\ Phys.\ Lett.}\ {\bf #1}\ (#2) #3}
\def\NC #1 #2 #3 {{\it Nuovo Cim.}\ {\bf #1} (#2) #3}
\def\NP #1 #2 #3 {{\it Nucl.\ Phys.}\ {\bf #1}\ (#2) #3}
\def\PL #1 #2 #3 {{\it Phys.\ Lett.}\ {\bf #1}\ (#2) #3}
\def\PR #1 #2 #3 {{\it Phys.\ Rev.}\ {\bf #1}\ (#2) #3}
\def\PP #1 #2 #3 {{\it Phys.\ Rep.}\ {\bf #1}\ (#2) #3}
\def\PRL #1 #2 #3 {{\it Phys.\ Rev.\ Lett.}\ {\bf #1}\ (#2) #3}
\def\RMP #1 #2 #3 {{\it Rev.\ Mod.\ Phys.}\ {\bf #1}\ (#2) #3}
\def\CMP #1 #2 #3 {{\it Comm.\ Math.\ Phys.}\ {\bf #1}\ (#2) #3}
\def\ZP #1 #2 #3 {{\it Z.\ Phys.}\ {\bf #1}\ (#2) #3}
\def\E #1 #2 #3 {{\bf #1}\ (#2) #3 (E)}

\def\SM{$\cal{SM}$}

\begin{document}

\begin{titlepage}

\begin{flushright}
hep-ph/9609447\\
Freiburg--THEP 96/19\\
September 1996
\end{flushright}
\vspace{1.5cm}

\begin{center}
\large\bf
{\LARGE\bf 	Analytic evaluation of two--loop renormalization\\ 
		constants of enhanced electroweak strength\\[.2cm]
		in the Higgs sector of the Standard Model}

\vspace*{1cm}
\rm
{V. Borodulin}\\[.5cm]

{\em 	Institute for High Energy Physics}\\
{\em	Protvino, Moscow Region 142284, Russia}\\[1.5cm]
{G. Jikia}\\[.5cm]

{\em Albert--Ludwigs--Universit\"{a}t Freiburg,
           Fakult\"{a}t f\"{u}r Physik}\\
      {\em Hermann--Herder Str.3, D-79104 Freiburg, Germany}\\[1.5cm]
      
\end{center}
\normalsize

\begin{abstract}
We calculate renormalization constants $Z_{H,w}$, the Higgs and $W$--boson 
mass and tadpole counterterms in the on--mass-shell renormalization scheme to
two loops in the heavy--Higgs--boson limit $m_H\gg M_W$. Explicit analytic 
formulae are presented for the two--loop integrals with masses, which are not 
known in the literature. As an application, the analytic expression for the
two--loop leading correction to the fermionic Higgs boson width is obtained.
\end{abstract}

\vspace{3cm}

\end{titlepage}

\section{Introduction}

After the observation of the top quark signal at the Tevatron, 
the mechanism of the spontaneous electroweak symmetry breaking remains the
last untested property of the Standard Model (\SM{}).
Although the recent global fits to the precision electroweak data from LEP, 
SLC and Tevatron seem to indicate a preference to a light Higgs boson 
$m_H=149^{+148}_{-82}$~GeV, $m_H<450$~GeV ($95\%C.L.$) \cite{LEPEWWG},
it is definitely premature to exclude the heavy Higgs scenario. The reason is
that a restrictive upper bound for $m_H$ is dominated by the result on 
$A_{LR}$, which differs significantly from the \SM{} predictions \cite{ALR}. 
Without $A_{LR}$ upper bound on $M_H$ becomes larger than 600~GeV \cite{ALR},
which is not far in the logarithmic scale from the value of the order of 
1~TeV, where perturbation theory breaks down. In order to estimate the
region of applicability of the perturbation theory, the leading two--loop 
${\cal O}(g^4 m_H^4/M_W^4)$ electroweak corrections were under intense study 
recently. In particular, the high energy weak--boson scattering 
\cite{scattering}, corrections to the heavy Higgs line shape 
\cite{GhinculovvanderBij}, corrections to the partial widths of the Higgs 
boson 
decay to pairs of fermions \cite{fermi_G,fermi_DKR} and intermediate vector
bosons \cite{vector_G,vector_FKKR} at two--loops have been calculated.
All these calculations resort at least partly to numerical methods.  Even for
the two--loop renormalization constants in the Higgs scalar sector of the \SM{}
\cite{GhinculovvanderBij,MDR} complete analytical expressions are not known. 
In this paper we present our analytic results for these two--loop 
renormalization constants, evaluated in the on--mass--shell renormalization
scheme in terms of transcendental functions $\zeta(3)$ and the maximal value 
of the Clausen function $\mbox{Cl}(\pi/3)$.

\section{Lagrangian and renormalization}

The part of the \SM{} Lagrangian describing the Higgs scalar sector of the \SM{} 
in terms of the bare quantities is given by:
  
\begin{eqnarray}
&&{\cal L} = \frac{1}{2}\partial_\mu H_0\partial^\mu H_0
+ \frac{1}{2}\partial_\mu z_0\partial^\mu z_0
+ \partial_\mu w^+_0\partial^\mu w^-_0 
\nonumber\\
&& -\, \frac{{m_H^2}_0}{2v_0^2}\left(w_0^+w_0^- + \frac{1}{2}z_0^2
+ \frac{1}{2}H_0^2 + v_0 H_0 + \frac{1}{2}\delta v^2
\right)^2.
\label{lagrangian}
\end{eqnarray}
Here the tadpole counterterm $\delta v^2$ is chosen in such a way, that
a Higgs field vacuum expectation value is equal to $v_0$
\begin{equation}
v_0 = \frac{2\, {M_W}_0}{g}
\end{equation}
to all orders.

Renormalized fields are given by
\begin{equation}
H_0 = \sqrt{Z_H}\,H, \quad z_0 = \sqrt{Z_z}\,z, \quad 
w_0 = \sqrt{Z_w}\,w.
\end{equation}

At two--loop approximation the wave function renormalization constants,
tadpole and mass counterterms take the form

\begin{eqnarray}
\sqrt{Z} &=& 1 + \frac{g^2}{16\pi^2}\,\delta Z^{(1)}
+ \frac{g^4}{(16\pi^2)^2}\,\delta Z^{(2)};
\nonumber \\
\delta v^2 &=& \frac{1}{16\pi^2}\,{\delta v^2}^{(1)} 
+ \frac{g^2}{(16\pi^2)^2}\, {\delta v^2}^{(2)};
\label{constants} \\ 
{M_W}_{0} &=& M_W + \frac{g^2}{16\pi^2}\, \delta M_W^{(1)}
+ \frac{g^4}{(16\pi^2)^2}\, \delta M_W^{(2)};
\nonumber \\
{m_H^2}_{0} &=& m_H^2 + \frac{g^2}{16\pi^2}\, {\delta m_H^2}^{(1)}
+ \frac{g^4}{(16\pi^2)^2}\, {\delta m_H^2}^{(2)}.
\nonumber 
\end{eqnarray}
Since weak coupling constant $g$ is not renormalized at the leading order in 
$m_H^2$, the $W$--boson mass counterterm is related to the Nambu--Goldstone 
wave function renormalization constant by the Ward identity  ${M_W}_0=Z_w M_W$.

In the on--mass--shell renormalization scheme all the counterterms are fixed 
uniquely by the requirement that the pole position of the Higgs  and 
$W$--boson 
propagators coincide with their physical masses and the residue of the
Higgs boson pole is normalized to unity.

The one--loop counterterms equivalent to those used in 
\cite{GhinculovvanderBij,MDR} are given by

\begin{eqnarray}
{\delta v^2}^{(1)}&=&
m_H^{2}{\xi^\epsilon_H}\,\Biggl\{-{\frac {6}{\epsilon}}+3
-\, \epsilon \biggl(\frac{3}{2}+\frac{\pi^2}{8}\biggr)
\Biggr\};
\nonumber
\\
\frac{{\delta m_H^2}^{(1)}}{m_H^2}&=&
\frac{m_H^{2}{\xi^\epsilon_H}}{M_W^2}
\,\Biggl\{-{\frac {3}{\epsilon}}+3-{\frac {3\,\pi\,\sqrt {3}}{8}}
+\, \epsilon \biggl(-3+\frac{\pi^2}{16}+\frac{3\pi\sqrt{3}}{8}
+\frac{3\sqrt{3}C}{4}-\frac{3\pi\sqrt{3}\log{3}}{16}\biggr)
\Biggr\};
\nonumber
\\
\delta Z_H^{(1)}&=&{\frac {m_H^{2}{\xi^\epsilon_H}\,}{M_W^{2}}}
\Biggl\{\frac{3}{4}-{\frac {\pi\,\sqrt {3}}{8}}
+\, \epsilon \biggl(-\frac{3}{4}+\frac{3\pi\sqrt{3}}{32}
+\frac{\sqrt{3}C}{4}-\frac{\pi\sqrt{3}\log{3}}{16}\biggr)
\Biggr\};
\label{zh}
\\
\frac{\delta M_W^{(1)}}{M_W}&=&\delta Z_w^{(1)}=
-{\frac {m_H^{2}{\xi^\epsilon_H}}{16\,M_W^{2}}}\Biggl\{1
-\frac{3}{4}\epsilon\Biggr\}.
\nonumber
\end{eqnarray}
Here the dimension of space--time is taken to be $d=4+\epsilon$ and 
\begin{equation}
\xi^\epsilon_H=e^{\gamma\,\epsilon/2}\,
\left(\frac{m_H}{2\,\pi}\right)^\epsilon.
\end{equation}
In contrast to papers \cite{GhinculovvanderBij,MDR} we prefer not to keep the 
one--loop
counterterms of ${\cal O}(\epsilon)$ order, {\it i.e.} unlike in the 
conventional on--mass--shell scheme used in \cite{GhinculovvanderBij,MDR},
we require that the one-loop normalization conditions are fulfilled only in 
the limit $\epsilon\to 0$, where the counterterms of the order 
${\cal O}(\epsilon)$ do not contribute. Such a modified on--mass-shell scheme
is equally consistent as the conventional one or 
the standard scheme of minimal dimensional renormalization, which assumes 
only the subtraction of pole terms at $\epsilon=0$ \cite{BR}. 
(Moreover, in general one cannot subtract all the nonsingular 
${\cal O}(\epsilon)$
terms in the Laurent expansion in $\epsilon$, as they are not polynomial
in external momenta.) The ${\cal O}(\epsilon)$ one--loop counterterms 
considered in \cite{GhinculovvanderBij,MDR} do can really combine with the 
$1/\epsilon$ terms at two--loop order to give finite contributions, but these
contributions are completely canceled by the additional finite parts of the 
two--loop counterterms, fixed through the renormalization conditions in the 
on--mass--shell renormalization scheme. 
The reason is that after the inclusion of the one--loop counterterms all the 
subdivergences  are canceled and only the overall divergence remains, which  
is to be canceled by the two--loop counterterms. The account of finite 
contributions coming from the combination of  ${\cal O}(\epsilon)$ one--loop 
counterterms  with $1/\epsilon$ overall divergence just redefines the finite 
parts of the two--loop counterterms. An obvious advantage of this modified
on--mass--shell scheme is that the lower loop counterterms once calculated 
could be directly used in higher loop calculations, while in the conventional
on--mass--shell scheme for $l$-loop calculation one needs to recalculate the 
one-loop counterterms to include all the terms up to 
${\cal O}(\epsilon^{l-1})$,
two-loop counterterms to include ${\cal O}(\epsilon^{l-2})$ terms and so on.

\section{Analytic integration}

The calculation of the Higgs and $W$--boson (or Nambu--Goldstone $w$, $z$ 
bosons) two--loop self energies, needed to evaluate
the renormalization constants (\ref{constants}), reduces to the evaluation
of the two--loop massive scalar integrals
\begin{eqnarray}
&&J(k^2;
n_1\, m_1^2,n_2\, m_2^2,n_3\, m_3^2,n_4\, m_4^2,n_5\, m_5^2) 
= -\frac{1}{\pi^4}\int\,D^{(d)}P\,D^{(d)}Q  \,\biggl(P^2-m_1^2\biggr)^{-n_1}\\
&& \times 
\biggl((P+k)^2-m_2^2\biggr)^{-n_2}
\biggl((Q+k)^2-m_3^2\biggr)^{-n_3}\biggl(Q^2-m_4^2\biggr)^{-n_4}
\biggl((P-Q)^2-m_5^2\biggr)^{-n_5}
\nonumber
\end{eqnarray}
and their derivatives $J'$ at $k^2=m_H^2$ or at $k^2=0$.

The most difficult is a calculation of the all--massive scalar master
integral corresponding to the topology shown in the Fig.~1. 

\vspace*{0.4cm}
\setlength{\unitlength}{1cm}
\begin{picture}(15,3)
\put(5,0){\epsfig{file=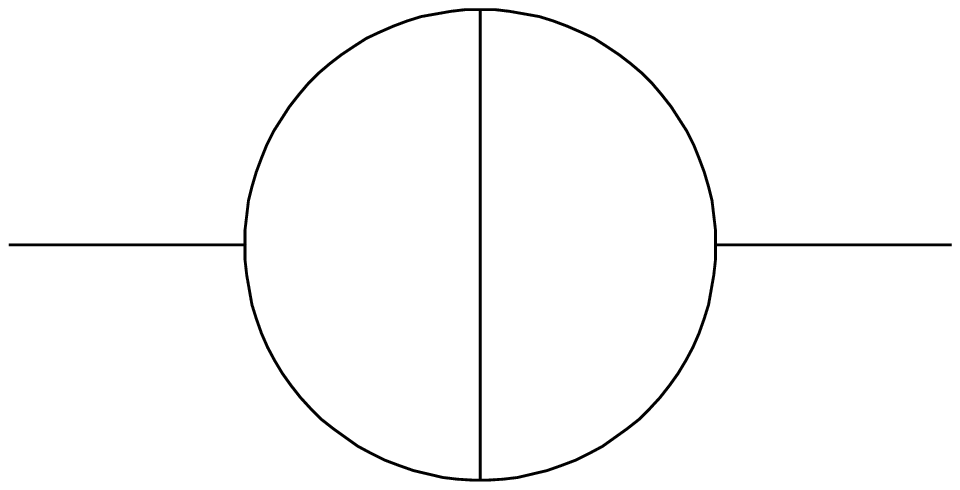,height=3cm}}
\end{picture}

\begin{center}
\parbox{6in}{\small\baselineskip=12pt Fig.~1. 
The two loop all--massive master graph. Solid line represents Higgs bosons. 
}
\end{center}
\vspace*{0.4cm}

This integral has a discontinuity that is an elliptic integral, resulting from
integration over the phase space of three massive particles, and is not 
expressible in terms of polylogarithms. However one can show 
\cite{ScharfTausk} that on--shell $k^2=m_H^2$ or at the threshold 
$k^2=9 m_H^2$ this is not the case. We use the dispersive method 
\cite{Broadhurst,BaubergerBohm} to evaluate
this finite integral on the mass shell:
\begin{equation}
m_H^2\, J(k^2;m_H^2,m_H^2,m_H^2,m_H^2,m_H^2)=
\sigma_a(k^2/m_H^2)+\sigma_b(k^2/m_H^2),
\label{master}
\end{equation}
where $\sigma_{a,b}$ correspond to the dispersive integrals 
calculated, respectively, from the two-- and three--particle discontinuities, 
which are itself reduced to one--dimensional integrals. 
The $\tanh^{-1}$ functions entering $\sigma_{a,b}$ can be removed 
integrating by parts either in the dispersive integral \cite{Broadhurst},
or in the discontinuity integral \cite{BaubergerBohm}. By interchanging the
order of integrations the latter representation  gives the three--particle cut
contribution $\sigma_b$ as a single integral of logarithmic functions 
\cite{BaubergerBohm}. After some rather heavy analysis we obtain at 
$k^2=m_H^2$: 

\begin{eqnarray}
\sigma_a(1)&=&\int_0^1 dy \, \frac{8}{y^{4}-y^{2}+1}
\log \left({\frac {\left (y^{2}+1\right )^{2}}{y^{4}+y^{2}+1}}\right)
\left [{\frac {\left (y^{4}-1\right )\log (y)}{y}}
-{\frac {\pi\,y}{\sqrt {3}}}\right ]  \nonumber \\
&=&{\frac {17}{18}}\,\zeta(3)-{\frac {10}{9}}\,\pi \,C+\pi ^{2}\,\log {2}
-{\frac {4}{9}}\,\pi ^{2}\,\log {3},
\\
\sigma_b(1)&=&\int_0^1 dy \, 2\,
\log \left({\frac {y^{2}+y+1}{y}}\right) \nonumber \\
&&\left [{\frac {\log (y+1)}{y}}
+{\frac {{\frac {\pi\,}{\sqrt {3}}}\left (y^{2}-3\,y+1\right )
-2\,\left (y^{2}-1\right )
\log (y)}{y^{4}+y^{2}+1}}-{\frac {\log (y)}{y+1}}\right ] \nonumber\\
&=&{\frac {1}{18}}\zeta(3)+{\frac {4}{9}}\,\pi \,C-\pi ^{2}\,\log {2}
+{\frac {4}{9}}\pi ^{2}\,\log {3}.
\end{eqnarray}
Here
$
C = \mbox{Cl}(\pi/3) = \im \,\mbox{li}_2\left(\exp\left(i\pi/3\right)\right)= 
1.01494\: 16064\: 09653\: 62502\dots
$

As a result we find 
\begin{eqnarray}
m_H^2\, J(m_H^2;m_H^2,m_H^2,m_H^2,m_H^2,m_H^2)&=&
\zeta(3)-{\frac {2}{3}}\,\pi\,C \nonumber \\
&=&-0.92363\: 18265\: 19866\: 53695 \dots
\label{HHHHH}
\end{eqnarray}
The numerical value is in agreement with the one, calculated using the 
momentum expansion \cite{DavydychevTausk}, and with the numerical values 
given in \cite{MDR,Adrian}.

Given the value (\ref{HHHHH}), the simplest way to calculate the derivative 
of the integral (\ref{master}) is to use Kotikov's method of differential 
equations \cite{Kotikov,MDR}
\begin{eqnarray}
m_H^4\, J'(m_H^2;m_H^2,m_H^2,m_H^2,m_H^2,m_H^2)&=&
{\frac {2}{3}}\,\pi\,C-\zeta(3)-{\frac {\pi^{2}}{9}}
\nonumber \\
&=&-0.17299\: 08847\: 12284\: 42069 \dots
\end{eqnarray}
 
All the other two--loop self energy scalar integrals contain ``light'' 
particles (Nambu--Goldstone or $W$, $Z$--bosons) and some of them are
IR divergent in the limit $M_{W,Z}\to 0$. In principle, one can 
calculate these IR divergent integrals in Landau gauge, where masses
of Nambu--Goldstone bosons are equal to zero and IR divergences are
represented as (double) poles at $\epsilon=0$ \cite{MDR}. However, in order to
have an additional check of the cancellation in the final answer of all the
IR divergent $\log(M_{W,Z}^2)$--terms, we work in 't~Hooft--Feynman gauge.
For the infra--red finite integrals the correct answer in the leading order in
$m_H^2$ is obtained just by    
setting $M_{W,Z}=0$. We agree with the results for these integrals given in
\cite{MDR,ScharfTausk}. The two--loop IR divergent integrals correspond
to the topologies shown in the Fig.~2, which contain ``massless'' 
propagators squared.

\vspace*{0.4cm}
\setlength{\unitlength}{1cm}
\begin{picture}(15,3)
\put(1,0){\epsfig{file=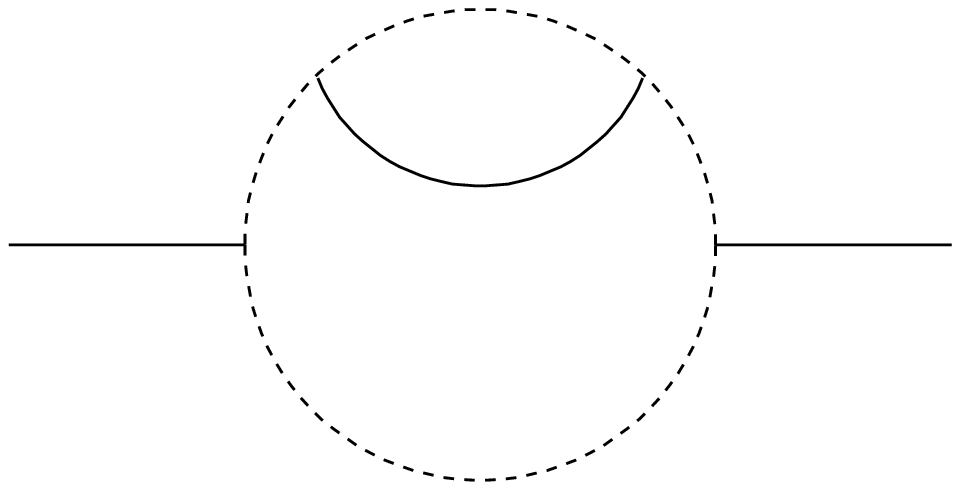,height=3cm}}
\put(10,0){\epsfig{file=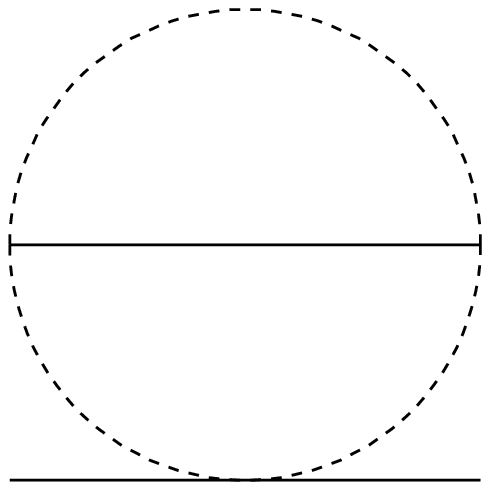,height=3cm}}
\end{picture}

\begin{center}
\parbox{6in}{\small\baselineskip=12pt Fig.~2. 
The two--loop IR divergent graphs. Dashed line represents ``light'' 
particles.}
\end{center}
\vspace*{0.4cm}

The relevant technique to handle these diagrams follows from the so--called
asymptotic operation method \cite{Tkachov}.
According to the recipe of As--operation, the formal Taylor
expansion in small mass $M_{W}$ entering propagator (or its powers) should 
be accomplished by adding the terms, containing the 
$\delta$--function or its derivatives.  The additional terms counterbalance
the infra--red singularities, arising in the formal expansion of propagator.
In our case we have  
\begin{eqnarray}
 \frac{1}{(P^{2} - M_{W}^{2})^{2}} & = & \frac{1}{(P^{2})^{2}}
+ 2\frac{M_{W}^{2}}{(P^{2})^{3}} + ...  \nonumber \\
&+& C_{1}(M_{W})\delta^{(d)}(P) + C_{2}(M_{W})\partial^{2}\delta^{(d)}(P)+ ...
\end{eqnarray}
Here the first coefficient functions $C_{i}(M_{W})$ read
\begin{eqnarray}
C_{1}(M_{W}) & = & \int D^{(d)}P \frac{1}{(P^{2} - M_{W}^{2})^{2}} 
\sim {\cal O}(M_{W}^{0}),  \\
C_{2}(M_{W}) & = & \frac{1}{2d}\int D^{(d)}P 
\frac{P^{2}}{(P^{2} - M_{W}^{2})^{2}} 
\sim {\cal O}(M_{W}^{2}). \nonumber 
\end{eqnarray}
This equality is to be understood in the following sense \cite{Tkachov}. 
One should integrate both parts of the equation multiplied by a test function
and then take the limit $d \rightarrow 4$. If one keeps in the expansion all 
terms up to order $M_{W}^{2 n}$, the resulting expression will represent a 
correct expansion of the initial integral to order $o(M_{W}^{2 n})$. 
To obtain the leading contribution to the diagrams Fig.~2,
it suffices just to take the first term of the Taylor expansion
and, correspondingly,  the first ``counterterm'' $C_{1}\delta^{(d)} (P)$.

Finally, the combination of Mellin--Barnes representation and Kotikov's 
method gives the following answer, corresponding to the first graph in 
Fig.~2 neglecting the terms of order ${\cal O}(M_W^2/m_H^2)$:
\begin{eqnarray}
m_H^2\, J(m_H^2;2\,M_W^2,M_W^2,0,M_W^2,m_H^2)&=&
{\xi^{2\epsilon}_H}\Biggl ({\frac {2\,i\,\pi}{\epsilon}}-{\frac {i\,\pi}{2}}
+\frac{2}{\epsilon}\,\log (\frac{M_W^2}{m_H^2})
\\
&&-\frac{1}{2}+{\frac {5\,\pi^{2}}{6}}
-\log (\frac{M_W^2}{m_H^2})
+\frac{1}{2}\log^{2} (\frac{M_W^2}{m_H^2})\Biggr );
\nonumber \\
m_H^4\, J'(m_H^2;2\,M_W^2,M_W^2,0,M_W^2,m_H^2)&=&
{\xi^{2\epsilon}_H}\Biggl (-{\frac {2\,i\,\pi}{\epsilon}}+2\,i\,\pi
-\frac{2}{\epsilon}\left(1\,+\,\log (\frac{M_W^2}{m_H^2})\right) \\
&&+2-{\frac {5\,\pi^{2}}{6}}
+\log (\frac{M_W^2}{m_H^2})
-\frac{1}{2}\log^{2} (\frac{M_W^2}{m_H^2})\Biggr ).
\nonumber 
\end{eqnarray}
The integral diverges as $1/\epsilon$, while if we would set $M_W=0$ from the
very beginning, it would diverge as $1/\epsilon^2$. The integral corresponding
to the second graph in the Fig.~2 up to ${\cal O}(M_W^2/m_H^2)$ is 
\begin{eqnarray}
J(m_H^2;2\,M_W^2,0,0,M_W^2,m_H^2)&=&
{\xi^{2\epsilon}_H}\Biggl ({\frac {2}{\epsilon^{2}}}
+\frac{1}{\epsilon}\left(2\,\log (\frac{M_W^2}{m_H^2})-1\right) \\
&&+\frac{1}{2}-{\frac {\pi^{2}}{12}}
-\log (\frac{M_W^2}{m_H^2})
+\frac{1}{2}\log^{2} (\frac{M_W^2}{m_H^2})\Biggr) \nonumber
\end{eqnarray}

The two--loop vacuum integrals needed to evaluate the tadpole counterterm 
${\delta v^2}^{(2)}$ have been calculated in \cite{vanderBijVeltman}.

\section{Results}

The analytic results for the two--loop renormalization constants are:
\begin{eqnarray}
\delta {v^2}^{(2)}&=&
\frac{m_H^{4}{\xi^{2\epsilon}_H}}{16\,M_W^2}
\left (
{\frac {72}{\epsilon^{2}}}
+{\frac {36\,\pi\,\sqrt {3}-84}{\epsilon}}
-162-3\,\pi^{2}+60\,\sqrt {3}C
\right );
\label{dv2}
\\
\frac{{\delta m_H^2}^{(2)}}{m_H^2}&=& 
\frac{m_H^{4}{\xi^{2\epsilon}_H}}{64\,M_W^4}\Biggl (
{\frac {576}{\epsilon^{2}}}
+\frac {144\,\pi\,\sqrt {3}-1014}{\epsilon}
\nonumber \\
&&+{\frac {99}{2}}-252\,\zeta(3)
+87\,\pi^{2}
-219\,\pi\,\sqrt {3}
\nonumber \\
&&+156\,\pi\,C
+204\,\sqrt {3}C
\Biggr);
\label{mh2}
\\
\delta Z_H^{(2)}&=&\frac{m_H^{4}{\xi^{2\epsilon}_H}}{64\,M_W^4}
\Biggl ( {\frac {3}{\epsilon}} - {\frac {75}{4}} - 126\,\zeta(3)
\nonumber \\
&&+{\frac {25\,\pi^{2}}{2}}-76\,\pi\,\sqrt {3}+78\,\pi\,C+108\,\sqrt {3}C
\Biggr );
\label{zh2}
\\
\frac{\delta M_W^{(2)}}{M_W}&=&\delta Z_w^{(2)}=
\frac{m_H^{4}{\xi^{2\epsilon}_H}}{64\,M_W^4}\left (
{\frac {3}{\epsilon}}-\frac{3}{8}-{\frac {\pi^{2}}{6}}+{\frac {3\,
\pi\,\sqrt {3}}{2}}-6\,\sqrt {3}C
\right ).
\label{zw2}
\end{eqnarray}
For comparison we have also calculated these counterterms following the
renormalization scheme \cite{GhinculovvanderBij,fermi_G} and keeping the 
${\cal O}(\epsilon)$ terms and found complete agreement with their (partly
numerical) results. In this scheme the renormalization constants 
(\ref{dv2}), (\ref{mh2}) look a bit more complicated due to the presence of 
the additional $\pi \sqrt{3}\log{3}$ terms
\begin{eqnarray}
\delta {v^2}^{(2)}&=&
\frac{m_H^{4}{\xi^{2\epsilon}_H}}{16\,M_W^2}
\Biggl (
{\frac {72}{\epsilon^{2}}}
+{\frac {36\,\pi\,\sqrt {3}-84}{\epsilon}}
+90-12\,\pi^{2}-12\,\sqrt {3}C
\\
&&-36\,\pi\,\sqrt{3}+18\,\pi\,\sqrt{3}\,\log{3}
\Biggr );
\label{dv2_e}\nonumber
\\
\frac{{\delta m_H^2}^{(2)}}{m_H^2}&=& 
\frac{m_H^{4}{\xi^{2\epsilon}_H}}{64\,M_W^4}\Biggl (
{\frac {576}{\epsilon^{2}}}
+\frac {144\,\pi\,\sqrt {3}-1014}{\epsilon}
\nonumber \\
&&+{\frac {2439}{2}}-252\,\zeta(3)
+63\,\pi^{2}
-363\,\pi\,\sqrt {3}
\nonumber \\
&&+156\,\pi\,C
-84\,\sqrt {3}C+72\,\pi\sqrt{3}\,\log{3}
\Biggr).
\label{mh2_e}
\end{eqnarray}
The wave function renormalization constants $Z_{H,w}$ are identical in these
two schemes.

As an example of the physical quantity, for which all the schemes should give
the same result, we consider the two--loop heavy Higgs correction to the
fermionic Higgs width \cite{fermi_G,fermi_DKR}. The correction is given by the
ratio 
\begin{eqnarray}
\frac{Z_H}{{M_W^2}_0/M_W^2}&=& 
1 + 2\frac{g^2}{16\,\pi^2}
\left(\delta Z_H^{(1)}-\frac{\delta M_W^{(1)}}{M_W} \right)
\nonumber \\
&&+ \frac{g^4}{(16\,\pi^2)^2}\Biggl[
2\,\frac{\delta M_W^{(1)}}{M_W}\,
\left (\frac{\delta M_W^{(1)}}{M_W}-\delta Z_H^{(1)}\right )
+\left (\delta Z_H^{(1)}-\frac{\delta M_W^{(1)}}{M_W}\right )^{2}
\nonumber \\
&&+2\,\delta Z_H^{(2)}-2\,\frac{\delta M_W^{(2)}}{M_W}
\Biggr].
\end{eqnarray}
Substituting (\ref{zh}), (\ref{zh2})--(\ref{zw2}) we find
\begin{eqnarray}
&&\frac{Z_H}{{M_W^2}_0/M_W^2}=
1\, +\, \frac{1}{8}
\frac{g^2}{16\, \pi^2}\frac{m_H^2}{M_W^2}\left( 13 - 2 \pi \sqrt{3}\right)
\\
&&+\, \frac{1}{16}\biggl(\frac{g^2}{16\, \pi^2}\frac{m_H^2}{ M_W^2}\biggr)^2
\left(3-63\,\zeta(3)-{\frac {169\,\pi\,\sqrt {3}}{4}}
+{\frac {85\,\pi^{2}}{12}}+39\,\pi\,C+57\,\sqrt {3}\,C\right).
\nonumber 
\end{eqnarray}
Again, we find complete agreement with the numerical result \cite{fermi_G}
and exact agreement with the result \cite{fermi_DKR}, taking into account
that their numeric constant $K_5$ is just minus our integral (\ref{HHHHH}).

\section*{Acknowledgments}

G.J. is grateful to J.J.~van~der~Bij and A.~Ghinculov for valuable 
discussions. This work was supported in part by the Alexander von Humboldt 
Foundation and the Russian Foundation for Basic Research grant 96-02-19-464.

\end{document}